# Simultaneously Determining Regional Heterogeneity and Connection Directionality from Neural Activity and Symmetric Connection

**Jiawen Chang[1,2], Zhuda Yang[1,2], Changsong Zhou[1,2,3*]**

1 Department of Physics, Hong Kong Baptist University, Kowloon Tong, Hong Kong

2 Centre for Nonlinear Studies, Hong Kong Baptist University, Kowloon Tong, Hong Kong

3 Life Science Imaging Centre, Hong Kong Baptist University, Kowloon Tong, Hong Kong

* Contact author: cszhou@hkbu.edu.hk

# Abstract:

The spatiotemporal patterns of neural dynamics are jointly shaped by directed structural interactions and heterogeneous intrinsic features of the neural components. Despite well-developed methods for estimating directionality in network connections from network of homogeneous nodes, how local heterogeneity impacts on directionality estimation remains poorly understood. In particular, the role of excitatory-inhibitory interactions in shaping network directionality and how these interactions should be incorporated into reconstruction frameworks remain largely unexplored. Here, we present a novel reconstruction framework that simultaneously estimates effective heterogeneity across network nodes and asymmetric network connections from neural activity and symmetric connection, both are assessible in experimental data, validated using macaque cortical connectivity data and several circuit models. We found that the estimated local heterogeneity remains consistent across various forms of parameterized local circuit heterogeneity. Furthermore, we demonstrated and quantified how hidden local inhibitory populations only modify within-region connection strengths, elucidating the functional equivalence between dynamics of excitatory-inhibitory networks and purely observing excitatory networks when estimating effective heterogeneity and asymmetry. Finally, we demonstrated the sampling interval effect in estimating network interactions with respect to the sampling resolution. Together, our results not only provide a unified framework for evaluating relative functional contributions of local heterogeneity and asymmetry to overall system dynamics but also reveal the fundamental limitations and scaling principles in reconstructing neural circuit connectivity from experimental observations.

## Author summary:

How heterogeneous brain regions communicate via directed connectivity to shape the neural dynamics patterns is a fundamental question in neuroscience. Traditional methods for estimating connectivity patterns from neural activity often assume all brain regions are homogeneous. However, how this regional heterogeneity due to anatomical difference impacts on directed connectivity estimation remains an open question. Here, we developed an approach that can simultaneously identify the direction of connectivity between regions and regional properties from existing brain activity data, which we validated using macaque brain connectivity data and different biological neurodynamic models. We found that our estimates of regional heterogeneity remain consistent across various types of circuit complexity. We further demonstrated robustness of this method when facing two key limitations: the inability to directly measure inhibitory neurons, revealing the functional equivalence between networks with and without inhibitory components and the effect of sampling resolution on network estimation.

## Introduction:

The biophysical large-scale dynamics of the whole brain cortex are shaped by inter-areal connections and the intrinsic local circuit properties of each brain region. At the microscale, neurons within neural circuits interact and entangle, forming neuronal populations with spatially distributed anatomical features, such as neurotransmitter receptor profiles [1-4], neuron density [5,6] and myelin content [7-9]. These features collectively contribute to the heterogeneity of brain regions at the macroscale [10,11]. Meanwhile, information transmission follows the fundamental rule that signals propagate via synapses from presynaptic to postsynaptic neurons, primarily through axons and dendrites [12,13]. Consequently, long-range white matter connections inherently reflect the directionality of true asymmetric structural connectivity (SC), which describes the neural interactions between cortical areas [14,15]. The intricate interplay between directional structural connectivity and local heterogeneity shapes the complex large-scale dynamics of the brain, emphasizing the importance of considering both factors to more comprehensively understand neural information flow and processing [16-19].

Developments in magnetic resonance imaging (MRI) have provided a noninvasive method to measure human brain heterogeneity and connectivity, both structural and functional, at the whole brain level. *In vivo* studies have found and revealed abundant anatomical heterogeneity content, e.g., contrast ratio of T1- to T2-weighted maps can reflect the intracortical myelination [7-9,20]. Information of SC is obtained through diffusion MRI (dMRI) and tractography methods, which estimate the density of white matter fibers connecting different brain regions [21-23]. However, SC obtained from dMRI does not contain directionality of the connection, which strongly limit our understanding of information processing in the brain network. Functional connectivity (FC) measures the temporal correlations between neural activity in these regions, typically calculated using resting-state functional magnetic resonance imaging (fMRI) data which captures blood oxygen level-dependent (BOLD) signals to represent neural synchronization dependence, again without the directionality of interaction [24]. Furthermore, these methods are not well-suited in quantifying the precise contribution of anatomical heterogeneity to brain dynamics and function, especially its interplay with the directionality of structural connectivity underlying information transmission [25].

Inferred from observed neural activity under assumed generative models, causal influences between cortical areas provide valuable insights into network interactions [26]. Various methods have been developed to estimate these influences, often referred to as effective connectivity (EC): Dynamic Causal Modeling, based on a Bayesian framework, primarily estimates latent neural quantities from measured brain activity [27-30]; perturbation studies systematically alter neural activity in specific nodes to map causal influences and information flow [31-33]; and noise-correlation analysis approaches establish relationships between SC and statistical quantities of neural activity (e.g., FC, covariance, and differential covariance) [34,35]. These methods collectively aim to capture the directional nature of information flow in brain networks, revealing causal relationships between regions of interest (ROIs) that go beyond simple activity correlations. However, these approaches often do not adequately account for local heterogeneity, which is also a key aspect contributing to dynamics and information processing [36-39], potentially resulting in estimated directional EC that may confound local heterogeneity with true directional connectivity. It is thus important to develop methods that can simultaneously estimate both local heterogeneity and connection asymmetry to more accurately characterize the directionality of connections in neural networks.

In this study, we propose a framework to simultaneously reconstruct node heterogeneity and asymmetry connections of brain networks, by extending the existing Dynamical Differential Covariance (DDC) method [35] initially developed for homogeneous nodes. We evaluated the reconstruction performance across a wide range of parameters on well-developed large-scale circuit models constrained by ground truth asymmetric macaque cortical connectivity and regional heterogeneity [36,37,40,41]. Firstly, this method systematically identifies effective local heterogeneity and asymmetric SC based on neural dynamics and symmetric SC, which are both accessible in empirical data, without requiring prior knowledge of SC directionality. We found that the effective heterogeneity identified by this method can further reconstruct various types of parameterized local heterogeneity across different models, such as self-recurrent strength, external input current, time constant, and firing threshold, enabling comparisons of the dynamical properties associated with different forms of local heterogeneity. We then demonstrated that this reconstruction method can effectively estimate the mixed effects involving hidden local inhibitory populations within a detailed excitatory-inhibitory network activity into effective excitatory-excitatory interactions in the regime below bifurcation to oscillatory states. Furthermore, we demonstrated a sampling interval effect on the reconstruction and separation of heterogeneity and asymmetry, considering the mismatch between observed data sampling rates and underlying neural dynamics temporal resolution. Together, our results provide a theoretical framework for reconstructing brain heterogeneity and connection asymmetry, and suggest a unified expression for further analysis of the relative contributions of local heterogeneity and asymmetric connections to neural network dynamics. The framework is extendable to other dynamical networks beyond neural systems.

# Results:

## I. A Unified Framework of Regional Heterogeneity and Asymmetric Structural Connections

The EC of the nervous system, inferred from neural activity, not only describes the strength of interactions between areas or neurons but also reveals the direction of information flow from one to another [28-34]. This directionality emerges from two key factors: the anatomical asymmetry structure of connections and the intrinsic properties of brain regions. Regional heterogeneity quantifies spatial variation across areas, establishing each region's hierarchical position and processing specialization, while asymmetry measures directional imbalances between feedforward and feedback pathways, revealing connection strength patterns and network topology [10,15]. However, little is discussed regarding how these two factors -- anatomical direction of inter-areal interactions and the heterogeneous features within areal properties – jointly influence the directionality of EC.

In this section, we describe a unified framework for representing regional heterogeneity and asymmetric connections. As the choice of regional heterogeneity relies on the model, without loss of generality, we started by following existing studies on heterogeneous large-scale brain network model [35,40]. For each ROI in total of $N$ regions, the large-scale circuit model describes the neural activity in the following dynamics:

$$\frac{dS_i}{dt} = -\frac{S_i}{\tau_s} + \gamma(1 - S_i)H(x_i) + \sigma v_i(t),$$
(1)

$$H(x_i) = \frac{ax_i - b}{1 - \exp(-d(ax_i - b))},$$
(2)

$$x_i = w_iS_i + G\Sigma_j C_{ij}S_j + I_i,$$
(3)

where $S_i(t)$ is the synaptic gating variable of region $i$. $\tau_s$ are the kinetic parameters controlling the decay time and $\gamma$ is scaling factor. $v_i(t)$ is independent standard Gaussian noise term with amplitude $\sigma$ at each ROI. The population firing rate (or activation function) $H(x_i)$ of region $i$ is defined as a function of total input current $x_i$, with gain factor $a$, threshold $b$ and parameter $d$ controlling the nonlinearity [42].

We utilized an open dataset containing directed SC and anatomical heterogeneity across macaque cortical areas to establish ground truth asymmetric SC and parameter heterogeneity [10,15,43]. $C_{ij}$ is asymmetric SC matrix describing anatomical connection from region $j$ to $i$ in total $N$ regions, ranging from primary sensory cortex to higher order cortex [10,15] [Materials and Method]. Following previous studies [36,41], local recurrent strength $w_i$, and external input $I_i$ were assumed to be heterogeneous across different regions (Figure 3A). Local recurrent strength $w_i$ is rescaled from empirical anatomical heterogeneity [Materials and Method]. The values of parameters in Eqs. 1–3 are provided in Table 1 following previous studies [36,40-43].

**Table 1 Fixed Parameters for Large-scale Circuit Model.**

| Parameter | Value | Reference |
|---|---|---|
| $\tau_S$ | 0.1 $s$ | [40] |
| $\gamma$ | 0.641 | [40] |
| $\sigma$ | 0.01 | [40] |
| $a$ | 270 $nC^{-1}$ | [40] |
| $b$ | 108 Hz | [40] |
| $d$ | 0.154 $s$ | [40] |
| $w_i$ | 0.0652-0.1581 $nA$ | [36,43] |
| $I_i$ | 0.30-0.35 $nA$ | [36] |

To directly link how regional heterogeneity and asymmetry contribute to the whole network dynamics, we linearized the model of Eqs.1-3 by performing the first-order Taylor expansion around its stable states [40,45]. This results in the Jacobian matrix governing the linear neural dynamic [Materials and Method]:

$$\frac{dS}{dt} = J(S^*)(S - S^*) + \sigma v(t),  \tag{4}$$

where $J(S^*)$ is the Jacobian matrix around the fixed point $S^*$. In the linear case, this Jacobian matrix represents EC. The elements in the Jacobian matrix quantify how local neural activity dynamics emerge from the interplay between asymmetric SC and regional heterogeneity (defined in Eq. 19).

By evaluating this Jacobian around fixed points (Eq. 4), we explicitly relate anatomical connection strengths (SC weights) and node-specific properties (state-dependent effective heterogeneity) to functional interactions governing activity patterns. Here, we propose that the directionality of the Jacobian $J$ (i.e. $J(S^*)$) can be expressed in the generic form of Eq. 5 below, with contribution from regional effective heterogeneity $h_i$ and asymmetric SC $C_{ij}$ between two regions:

$$J_{ij} = h_i C_{ij}, \qquad i \neq j, \tag{5}$$

and in the case of the model in Eqs. 1-3,

$$h_i = \gamma G (1 - S_i^*) \frac{\partial H}{\partial x_i} |_{x_i = x_i^*}, \tag{6}$$

and the diagonal elements of the Jacobian $J_{ii} = -\frac{1}{\tau_S(1 - S_i^*)} + \frac{w_i h_i}{G}$ [Materials and Method]. $x_i^* = w_i S_i^* + G \Sigma_j C_{ij} S_j^* + I_i$ is the steady total input.

This formula for both representing effective regional heterogeneity and asymmetric connections described by Eq. 5 is not restricted to a specific dynamical model but is a unified framework that can encompass a broad range of heterogeneous dynamic models [36-38,41]. Table 2 presents the expression of effective heterogeneity $h_i$ across different models, as discussed in the subsequent reconstruction procedure.

**Table 2 Expressions of Effective Heterogeneity $h_i$ Across Different Models.**

| Models | Reference | Population(s) | $h_i$ |
|--------|-----------|---------------|-------|
| $A$ | Kong et al. (2021) [36] | $E$ | $\gamma G(1-S_i^*)\frac{\partial H}{\partial x_i}\vert_{x_i=x_i^*}$ |
| $B$ | Adapted from Deco et al. (2021) [37] | $E$ | $\gamma G(1-S_i^*)\frac{\partial H_i}{\partial x_i}\vert_{x_i=x_i'}$ |
| $C$ | Demirtas et al. (2019) [38] | $E, I$ | $(E): \gamma G\left(1-S_{i,E}^*\right)\frac{\partial H_E}{\partial x_i}\vert_{x_{i,E}=x_{i,E}^*}$ |

In all models, $S_i^*$ and $x_i^*$ represent the steady state solution and steady overall external current of region $i$ respectively. $x_i'$ and $x_{i,E}^*$ have different notations because their formula are different from Model A. In Model B, the effective heterogeneity $h_i$ is determined by both the heterogeneous activation function $H_i$ and the overall input $x_i'$ [Materials and Method]. Model B combines various parameterized local heterogeneities $\{\tau_i, b_i\}$ in contrast to Model A which has $\{w_i, I_i\}$ (see Section V). In Model C, the effective heterogeneity $h_i$ is determined by both the parameterized local heterogeneities and the interactions between excitatory and inhibitory populations (see Section V).

The following sections systematically demonstrate and validate our framework. We first show how to reconstruct the effective heterogeneity $h_i$ and asymmetric SC $C_{ij}$ from the neural activity $S_i(t)$ and symmetric SC $W_{ij}$ in section II, where we assume $W_{ij} = \frac{C_{ij}+C_{ji}}{2}$ and can be obtained from dMRI. We also evaluate the accuracy of our reconstruction method based on Model A (Eqs. 1-3) under different parameter conditions in section III.

Our proposition above demonstrates that the reconstructed effective heterogeneity $h_i$ is typically influenced by both the system's steady state $S^*$, the firing rate change (gain function level) $\frac{\partial H}{\partial x_i}\big|_{x_i=x_i^*}$ , as well as the parameterized anatomical heterogeneity, including the local recurrent strength $w_i$ and external input $I_i$.  We discuss how to use the effective heterogeneity $h_i$ to reconstruct the parameterized anatomical heterogeneity $w_i$ and $I_i$ in section IV. In section V, we present how our proposed effective heterogeneity can be derived from different combinations of heterogeneous parameters, representing a unified framework: when studying EC, effective heterogeneity characterizes the dynamical contributions of heterogeneity based on different empirical data sources. Finally, in section VI, we propose the sampling interval effect, systematically quantifying how the true structural connectivity and effective heterogeneity are affected by the sampling window during the estimation of EC under sampled neural activity.

## II. Neural Activity-Driven Reconstruction Method of Regional Heterogeneity and Asymmetric Connection

We developed a data-driven method to disentangle and identify two key components: the underlying effective local heterogeneity $h_i$ and asymmetric SC $C_{ij}$ following the generic framework in Eq. 4 and Eq. 5. This method requires only neural activity data and symmetric SC information, both of which are commonly available from whole-brain imaging studies. Briefly, our heterogeneity and asymmetry reconstruction framework consist of two parts: temporal reconstruction and spatial reconstruction (Figure 1). For the temporal reconstruction, we estimated the Jacobian matrix $J$ from the neural activity $S$, and for the spatial reconstruction, we inferred the regional heterogeneity $h$ and asymmetric SC $C$ based on estimated $J$ and symmetric SC $W = \frac{C+C^T}{2}$. We assumed that $W$ can be regarded as connectivity obtained in dMRI [21-23].

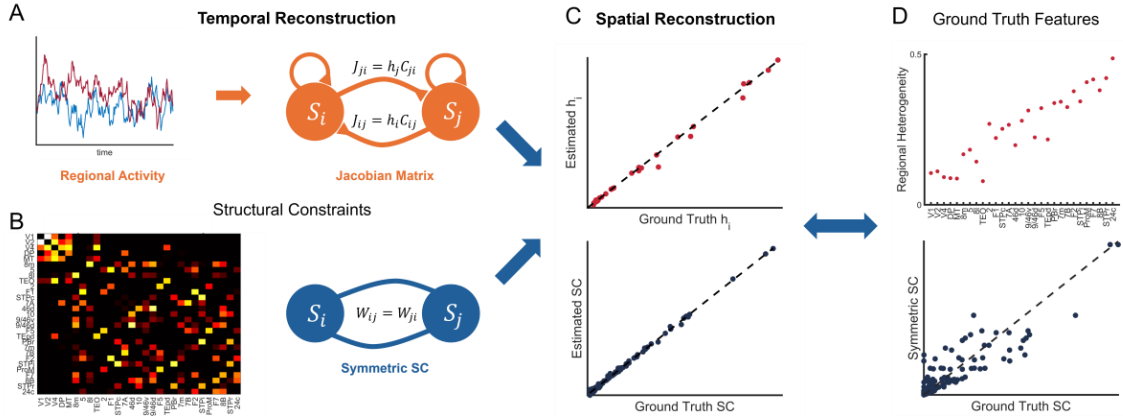

**Figure 1. Schematic diagram of reconstruction.** (A) Temporal Reconstruction. Neural activity $S(t)$ is transformed into a Jacobian matrix via Dynamical Differential Covariance (DDC) which provides an unbiased estimation of network coupling Jacobian $J_{ij}$. The Jacobian matrix (EC) here can be further divided into effective heterogeneity $h_i$ and asymmetric SC $C_{ij}$. (B) Symmetric SC $W_{ij} = \frac{C_{ij}+C_{ji}}{2}$, is considered as structural constraints for spatial reconstruction. (C) Spatial reconstruction further separates effective heterogeneity $h_i$ (top) and asymmetric SC $C_{ij}$ (bottom) following the temporal reconstruction and structural constraints. (D) The effective heterogeneity $h_i$ represents an example calculated using Eq.6. (Bottom) The systematic difference between ground truth asymmetric SC and symmetric SC approximation, with asymmetry level $\eta(C) = 0.7$, representing that this empirical SC is symmetry-dominated but regulated by asymmetric connections. $\eta(C)$ is calculated as the element-wise correlation between the upper and lower triangular matrices.

During temporal reconstruction (Figure 1A), we employed the DDC method to decode the directionality from neural activity data based on Eq. 4 and infer the underlying Jacobian $J$ representing the spatial network structure [35]. The DDC method provides an unbiased estimation of the directionality of Jacobian matrix by assuming that the observed neural activity arises from fluctuations around the stable fixed point in a noise-driven linear dynamical system in the form of Eq. 4. This approach derives EC through least square minimization, making it both computationally efficient and robust to noise [Materials and Method]:

$$\hat{J} = <\frac{dS}{dt}, S> <S, S>^{-1}. \tag{7}$$

We then employed spatial reconstruction with an empirical symmetric SC constraint to disentangle the effective heterogeneity $h_i$ and asymmetric SC $C_{ij}$ from the estimated Jacobian $\hat{J}_{ij}$ (Figure 1B and 1C). This constraint aligns with the properties of the acquired data, as structural connectivity derived from dMRI is inherently symmetric [21-23]. We define symmetric SC as the symmetric counterpart of the asymmetric SC: $W = \frac{C + C^T}{2}$. This results in the following overdetermined equations (for sufficiently large $N$):

$$\begin{cases} h_i C_{ij} = \hat{J}_{ij} \\ \frac{C_{ij} + C_{ji}}{2} = W_{ij} \end{cases}, i \neq j. \tag{8}$$

Let $y_i = \frac{1}{h_i}$, we can reshape Eq.8 into the form of a multivariate linear regression to estimate $\hat{y}_i$ [Materials and Method], then we can estimate the effective heterogeneity and asymmetric SC:

$$\begin{cases} \hat{h}_i = \dfrac{1}{\hat{y}_i} \\ \hat{C}_{ij} = \hat{y}_i \hat{J}_{ij} \end{cases}.$$ (9)

The estimation of the Jacobian $\hat{J}$ provides information about the network directionality inferred from neural activity, whereas the symmetric structural connectivity $W$ indicates the presence and strength of connections.

In summary, temporal reconstruction (Eq. 7) decodes dynamical directionality, while spatial reconstruction (Eq. 9) further disentangles anatomical asymmetry and heterogeneity. This method combines observed neural activity with empirical symmetric connectivity to simultaneously recover both asymmetric structural connectivity and effective regional heterogeneity.

## III. Robustness of Reconstruction in Parameter Exploration

To evaluate the accuracy of our reconstruction method in distinguishing and revealing connectivity asymmetry and regional heterogeneity, we applied it to Model A (Eqs. 1–3), which biophysically describes resting-state activity in cortical areas. Motivated by previous studies that employ global coupling strength $G$ to tune the dynamical regime for optimal empirical fits [36-38,40,41] and noise strength $\sigma$ to test the noise tolerance [35], we explored different values of $G$ and $\sigma$. For each parameter condition, we simulated the neural activity and calculated the ground truth Jacobian matrix $J_{ij}$ and effective heterogeneity $h_i$ according to Eqs. 19 and 6 [Materials and Method]. The simulated neural activity and symmetric SC were then used for the reconstruction procedure and validation as described in Figure 1A and 1C. The estimated features—Jacobian matrix $\hat{J}_{ij}$, asymmetric SC $\hat{C}_{ij}$, and effective heterogeneity $\hat{h}_i$ were then compared.

We evaluated the reconstruction performance for EC $J_{ij}$, asymmetric SC $C_{ij}$ and effective heterogeneity $h_i$ on model A by calculating the relative error (RE) between estimated and empirical features: $RE(a) = \frac{\|a_{emp} - a_{est}\|}{\|a_{emp}\|}$. We studied RE as a function of global coupling strength $G$, noise strength $\sigma$, and time length in neural activity (Figure 2 and Figure S2).

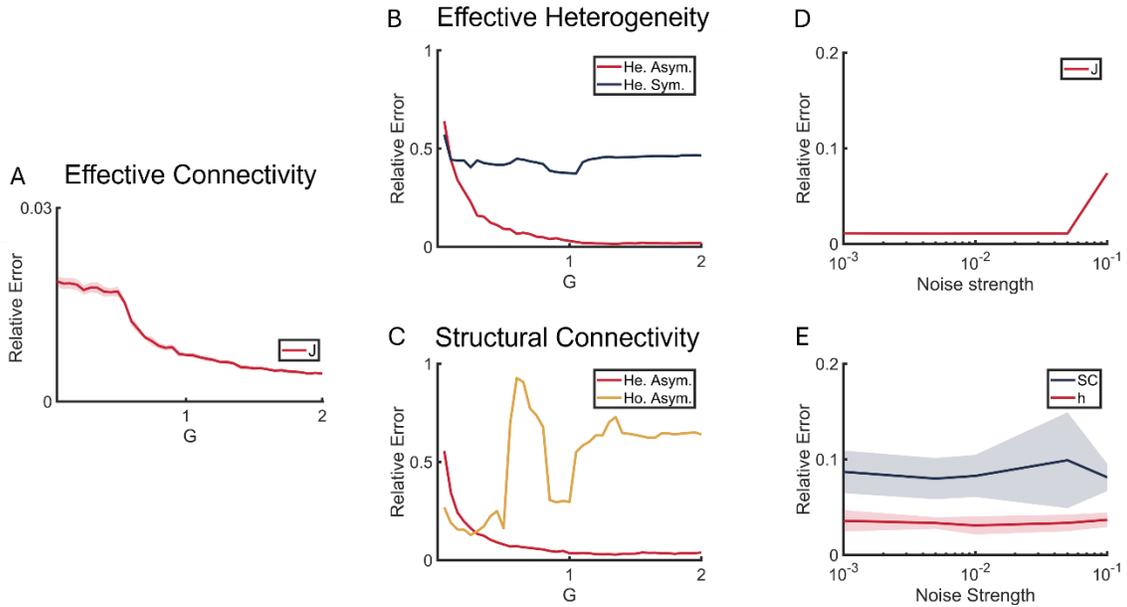

**Figure 2. Robustness of reconstruction in effective heterogeneity and asymmetry.** (A) The relative errors between ground truth and estimation of EC (Jacobian in Eq.9) as functions of the global coupling strength $G$. (B and C) The relative errors of effective heterogeneity $h_i$ (B) and asymmetric SC $C_{ij}$ (C) revealed by the reconstruction method as functions of $G$. Red line represents reconstruction considering both heterogeneity and asymmetry. Blue line is baseline comparison ignoring asymmetry (B). Yellow line is baseline comparison ignoring heterogeneity (C). (D) The relative errors of Jacobian as function of noise strength $\sigma$. (E) The relative errors of asymmetric SC $C_{ij}$ (blue) and regional heterogeneity $h_i$ (red) as functions of $\sigma$. The colored lines show the mean relative errors across 10 simulations, with shaded areas indicating one standard deviation from the mean. Simulation length $50,000s$.

The relative errors between the ground truth and the estimates of the Jacobian matrix $J_{ij}$, asymmetric SC $C_{ij}$ and effective heterogeneity $h_i$ decreased as the global coupling strength $G$ increased (Figure 2A-C, red lines). This behavior arises because higher $G$ corresponds to a more globally integrated state of the system, where neural activity inherently encodes more information about the underlying asymmetric connections. This can be explained with Eqs. 4 and 5 where the dynamic is controlled by the Jacobian matrix $J$ and the white noise $\eta$: larger $G$ can bring larger contribution to the activity fluctuation $\frac{dS}{dt}$ from the dynamics than the noise, reducing the noise effect in estimating $\hat{J}_{ij}$.

We systematically tested the consequences of ignoring key model features by comparing reconstruction performance under different assumptions. When estimating effective heterogeneity $h_i$ while ignoring ground truth asymmetry (namely assuming connectivity is symmetrical), baseline comparison shows consistently high relative errors around 0.5 (Figure 2B, blue line), whereas our full framework typically achieves <0.2 at non-localized states (G>0.3). Similarly, when estimating asymmetric SC while ignoring ground truth heterogeneity (namely assuming homogeneous nodes), baseline comparison shows higher relative errors at high levels of $G$ (Figure 2C, yellow line). We also note that this baseline relative error fluctuation is similar to the standard deviation (SD) of $h_i$ across regions (low SD indicates that $h_i$ values are nearly identical across regions, while high SD indicates substantial heterogeneity) (Figure S1C). This pattern suggests that systematic errors are strongly influenced by the relative heterogeneity level, while our framework demonstrates substantially better performance.

REs of EC (Jacobian), SC and effective heterogeneity as functions of noise strength $\sigma$ remained consistently low across the range of $\sigma$ but RE of EC slightly increased at $\sigma = 0.1$ (Figure 2C and 2D). This increase can also be explained by the fact that larger noise strength $\sigma$ will enlarge the bias in the activity fluctuation $\frac{dS}{dt}$ as in Eq. 4, and therefore degrade the estimation accuracy of EC (Eq. 7). Based on this, we selected $\sigma = 0.01$ as optimal value for our simulations to ensure the system is appropriately noise-driven [36,37].

A previous study of DDC has examined how data size affects estimation accuracy [35]. To evaluate this effect on our reconstruction performance, we compared the reconstructed features across different data lengths. The relative error decreased as the data length increased from 100s to 50,000s (Figure S1), confirming that longer recordings lead to more accurate reconstructions.

To further validate our framework's capability, we tested whether the framework can correctly identify the directionality of EC contributed by heterogeneity and/or asymmetry across various ground truth conditions, namely the ground truth models may or may not contain heterogeneity and/or asymmetry. Results demonstrated that our framework maintains comparable performance in recovering true effective heterogeneity and connectivity patterns as global coupling strength increases (Figure S2), validating its capacity to accurately distinguish between different sources of directional connectivity.

To assess method-independence, we compared our DDC-based approach with Lyapunov optimization (LO) for EC estimation [53]. LO uses gradient descent to solve for the EC that corresponds to observed covariance, based on the Lyapunov equation constraint in linear dynamics. Results confirm that our framework's performance is robust across different EC estimation methods, with DDC providing computational advantages particularly under high coupling conditions where the LO method faces convergence challenges (Figure S3).

Together, these results highlight the robustness of the reconstruction method across a broad range of model architecture, parameter spaces and data constraints, underscoring its potential for reliable inference under varying experimental conditions.

## IV. Detailed Reconstruction of Heterogeneous Parameters

In this section, we investigated the mapping relationship between estimated effective heterogeneity $h_i$ (Eq. 6) and specific parameterized local heterogeneity $w_i$ and $I_i$ in dynamical Model A (Figure 3A). Notably, compared to the diagonal elements of EC $J_{ii}$, the parameter $w_i$ represents regional self-connections while $I_i$ captures external inputs such as subcortical influences (Materials and Method, Eqs. 28 and 30). Estimating these two heterogeneity parameters provides more biologically meaningful interpretations of neural circuit dynamics.

We first evaluated the reconstruction performance of local recurrent strength $w_i$ and external input $I_i$ from effective heterogeneity $h_i$ in Model A as functions of global coupling strength $G$. The detailed reconstruction of local recurrent strength $\hat{w}_i$ was derived using $h_i$ and the diagonal elements of $\hat{J}$, and the external input $\hat{I}_i$ was computed by determining the fixed total input $x_i^*$ from firing rate change $\frac{dH}{dx_i}|x_i^*$ (Materials and Method, Eqs.28-30).

At low levels of $G$, the relative errors of both $w_i$ (blue) and $I_i$ (red) decreased as $G$ increased (Figure 3B), consistent with our findings from the spatial reconstruction of $h_i$ and $C_{ij}$ (Fig 2B). However, at higher values of $G$, the relative error of $I_i$ began to increase (Figure 3B, red).

It was reported that the best fit between model FC and empirical FC lies around the bifurcation point in the same model A [40]. To better understand how our reconstruction framework performs near this bifurcation, which represents a highly nonlinear regime, and to explain the poor reconstruction performance at both low and high levels of G, we examined three representative conditions: localized dynamics (where the dynamics only slightly departure from the quiescent fixed point at low global coupling strength $G = 0.1$, Figure 3B, red square); the first bifurcation (where the dynamics start to exhibit bistability at $G = 0.8$, blue circle); and the second bifurcation (where the dynamics bistability is about to vanish at global scale at $G = 1.3$, yellow triangle). Bistability was detected by simulating Model A at both low and high initial states to see whether each region would converge to distinct stable states across $G$ (Figure S4A).

Near the first bifurcation dynamics states, the firing rate change (gain function level) $\frac{dH}{dx_i}|_{x_i^*}$ were widely distributed within the linear range with some regions exhibited in the nonlinear regime (Figure 3C, blue circle).

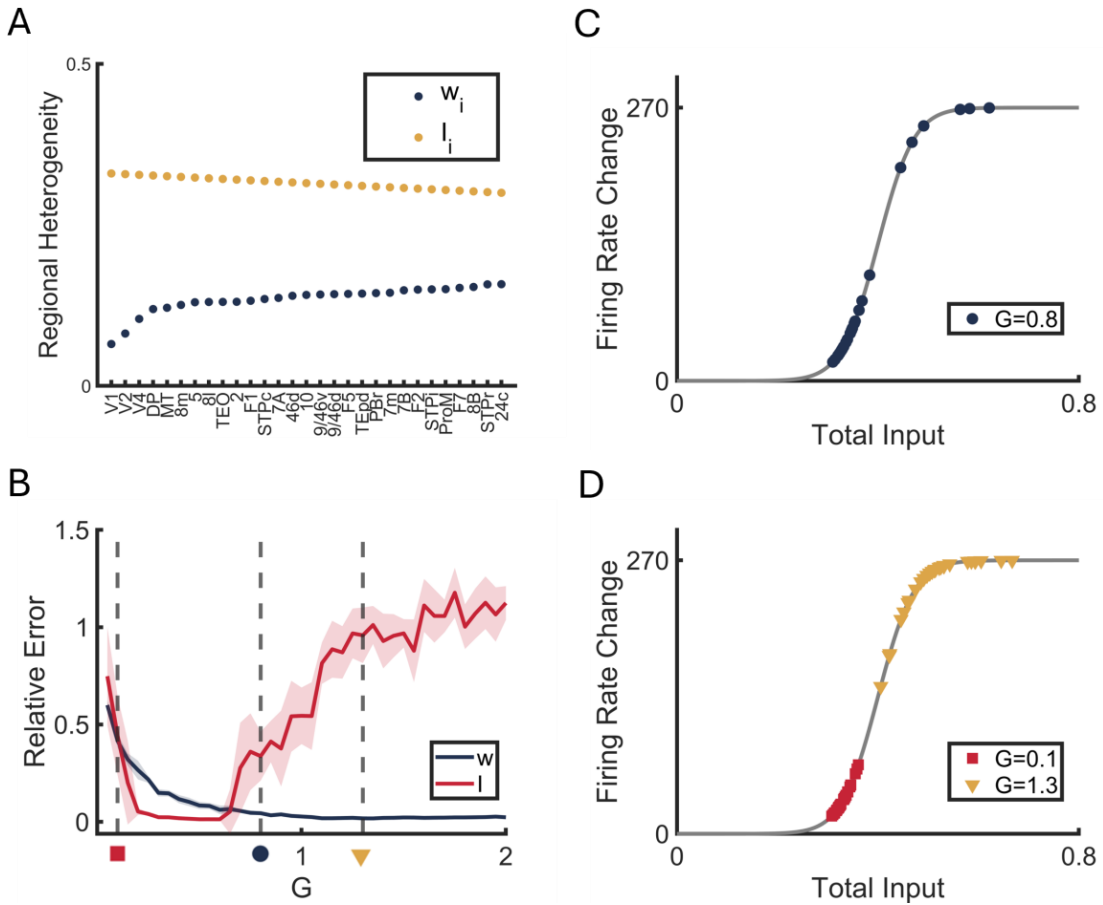

**Figure 3. Performance in reconstruction of detailed parameters.** (A) Ground truth features of Model A. The regional recurrent strength $w_i$ is rescaled from empirical anatomical heterogeneity [43], while the external input $I_i$ is set to decrease with hierarchy to be consistent with previous work [36]. (B) The relative errors in Model A between ground truth and estimation of $w_i$ (blue) and $I_i$ (red) as functions of $G$. Vertical dash lines represent different $G$ values: $G = 0.1$ (red square), $G = 0.8$ (blue round) and $G = 1.3$ (yellow triangle). (C and D) Derivative of firing rate $\frac{dH}{dx_i}(x_i)$ at stable states

$S^*$ for each ROI under three different global coupling strengths corresponding to B. Firing rate change of each ROIs under high global coupling are located at nonlinear region (D, yellow triangle). The colored lines show the mean value across 10 simulations, with shaded areas indicating one standard deviation from the mean. Simulation length $50,000s$.

However, in both extreme states at low and high levels of $G$, the firing rates are biased towards nonlinear regimes: near the second bifurcation, most ROIs operate in the upper saturation region (Figure 3D, yellow triangle), while in localized dynamics (red square), firing rates cluster near the lower threshold region. When operating in these nonlinear regions, small perturbations —such as noise in estimating the EC—may bias the estimation of $\hat{y}_i$ and $\frac{\partial H}{\partial x_i}|_{x_i^*}$. This can lead to large errors in reconstructing $x_i^*$, therefore introducing bias in the estimation of $\hat{I}_i$ due to the high sensitivity to noise in these regimes (Materials and Method, Eqs. 28 and 30).

Detailed regional relative errors are shown at three $G$ values (Figure S4C and D). We noted that with the increase of G, reconstruction performance of $w_i$ at all regions increased, while reconstruction performance of $I_i$ shows variability, consistent with previous studies that also reported inaccurate $I_i$ estimation [36].

To further examine whether the variability in reconstructed $\hat{I}_i$ significantly affects the dynamical system's ability to capture FC patterns, we re-simulated neural activity using the reconstructed parameters ($\hat{w}_i$, $\hat{I}_i$, and $\hat{C}_{ij}$) in Model A and compared the correlation between the re-simulated FC and the original ground truth FC. Our results demonstrate that even near the second bifurcation point ($G = 1.3$), the average correlation remains reasonable at 0.84, indicating that the inaccurate estimation of $I_i$ does not substantially impair the framework's capacity to reproduce essential FC characteristics (Figure S4B, Solid line). Notably, re-simulation using a linear model (Eq. S1) with estimated effective heterogeneity $\hat{h}_i$ and asymmetric SC $\hat{C}_{ij}$ successfully reproduces the FC patterns (Figure S4B, Dashed line). Together, these results suggest that our framework successfully captures the effective heterogeneity and asymmetric SC. Although detailed reconstruction of parameters suffers from nonlinearity, the reconstructed parameters can still reasonably reproduce the observed FC patterns.

## V Effective Heterogeneity Contributed by Different Heterogeneity and Unobserved Inhibitory Dynamics

Our analysis of Model A presented above demonstrates that effective heterogeneity is influenced not only by anatomical heterogeneity but also by the underlying dynamical states (Table 2). However, heterogeneous large-scale circuit modeling in macaque and human studies have incorporated various experimental data reflecting regional variations, emphasizing the variability in parameterizing empirical heterogeneity into the dynamical properties of each region [36,37,43]. Therefore, we propose that the effective heterogeneity identified through our EC separation framework provides a unified approach to characterize the dynamical contributions of heterogeneity based on different empirical data sources.

We next sought to investigate whether this framework for detailed reconstruction could be applied to an alternative large-scale heterogeneous model where regional heterogeneity occurs in other parameters, including effective heterogeneity and asymmetric SC reconstruction in E-I model.

We first utilized a modified dynamical mean-field model (Model B) [37] in which heterogeneity was introduced in other local parameters, specifically the timescale $\tau_i$ and firing threshold $b_i$, which govern the regional capacity for information loss and accumulation, while local recurrent strength $w$ and effective outer input $I$ were uniform across regions [Materials and Method, Eqs. 31-33]. The detailed reconstruction of timescale $\hat{\tau}_i$ and firing threshold $\hat{b}_i$ was computed using $\hat{h}_i$ and estimated EC $\hat{J}$.(Eqs.35 and 36).

We found that the ground truth $\tau_i$ and $b_i$ of Model B can be calculated from Eq. S2 using ground truth $w_i$ and $I_i$ of Model A, which provides a mapping relationship between these parameter pairs of the two models. The homogeneous $w$ and $I$ of Model B are the averaged of ground truth $w_i$ and $I_i$ of Model A.

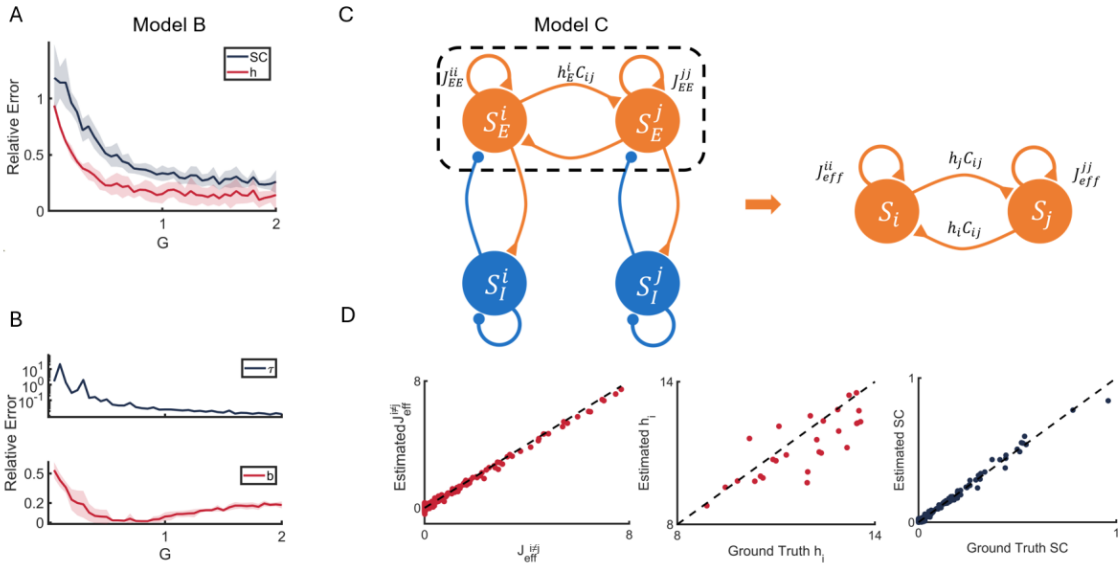

**Figure 4. Reconstruction Performance of Model B and Model C.** (A) The relative errors of reconstruction of the asymmetric SC (blue) and regional heterogeneity $h_i$ (red) of Model B as functions of $G$. (B) The relative errors between ground truth and estimation of $\tau_i$ (blue) and firing threshold $b_i$ (red) of Model B as functions of $G$. (C) Illustration of reconstruction procedure with localized excitation-inhibition interactions (Model C). The full model consists of an excitatory $S_E^i$ and an inhibitory population $S_I^i$ for each region ($i = 1, \ldots, N$), and only excitatory populations have inter-region connections: $h_E^i C_{ij}$ represents the directed connection from region $j$ to region $i$, with $C_{ij}$ denotes the SC and $h_E^i$ denotes the effective heterogeneity of excitatory population. $J_{EE}^{ii}$ represents the local recurrent strength of the excitatory population of region $i$. Dash square represents that only the activity of excitatory populations is observed for reconstruction. $J_{eff}^{ii}$ represents the effective local recurrent strength reconstructed from excitatory activity and $h_i$ is the effective heterogeneity absorbing the influence from inhibitory population. (D) Reconstruction example of Model C while only excitatory activity is observed. Ground truth $J_{eff}$ and $h_i$ are derived in Eqs. 11 and 13. Dash line represents $y = x$.

Similar to the results observed in Model A, the relative errors of both reconstructed asymmetric SC (Figure 4A, blue) and effective heterogeneity $h_i$ (Figure 4A, red) decreased as the global coupling strength $G$ increased. Furthermore, detailed reconstruction of the parameterized heterogeneity $\tau_i$ and $b_i$ showed good estimation performance within a moderate range of $G$ (Figure 4B). However, the relative reconstruction errors in Model B were higher than those in Model A, particularly at low values of $G$. This reveals that using $\tau_i$ as regional heterogeneity parameters may be more sensitive to noise.

We also demonstrated that the heterogeneous parameters of Model A and Model B exhibit similar dynamical properties on shaping the effective heterogeneity $h_i$ and localized dynamics (i.e., diagonal elements of the Jacobian matrices) and can be converted into one another. Specifically, in the spirit of dimensionless analysis, the heterogeneous $w_i$ and $I_i$ in Model A can be mapped onto the heterogeneous $\tau_i$ and $b_i$ in Model B while maintaining the same steady states and Jacobian matrix (Eq. S2). This was done by comparing the Jacobian matrices of each model (Eqs.27 and 34) at the same neural activity level $S^*$ [Materials and Method]. Consequently, the FC and autocorrelation timescales of the neural dynamics remain generally unchanged (see Supplementary Information, Figure S5). This is because in the steady state, the FC and autocorrelation are determined by the Jacobian obeying the Lyapunov equation [45]. Intuitively, the heterogeneity in $\tau_i$ and $b_i$ in Model B parallels the heterogeneity in $w_i$ and $I_i$ in Model A, as both sets of parameters control regional activity patterns. Variations in external input $I_i$ shift the baseline firing in the activation function $H$, similar to the role of firing thresholds $b_i$. Meanwhile, heterogeneous local recurrent strength $w_i$ modulates self-activation capability, functionally equivalent to the effect of decay time constants $\tau_i$. This validates our proposed framework, where the effective heterogeneity $h_i$ characterizes the dynamical contributions of different heterogeneous parameters and provides a top-down perspective on how to configure heterogeneous parameters to match empirically observed effective heterogeneity.

We next sought to extend our investigation by incorporating excitatory-inhibitory population dynamics to explore how inhibitory activity contributes to network reconstruction and modulates effective heterogeneity when only activity from the excitatory population is accessible. The balance between excitatory and inhibitory neural populations enriches the spatiotemporal patterns and stability of brain dynamics, underscoring the critical role of inhibitory populations in shaping sophisticated brain dynamics [43,46,47]. However, fluctuations of whole brain imaging data such as BOLD and magnetoencephalographic (MEG) signals are thought to be largely contributed by excitatory postsynaptic potentials rather than inhibitory postsynaptic potentials [44,48]. This physiological bias complicates the reconstruction of effective heterogeneity and connectivity asymmetry from such data.

Specifically, we analyzed the reconstruction performance of an excitatory-inhibitory mean-field model featuring a localized inhibitory population (Model C, Figure 4C). We consider the scenario where inhibitory activity is unobserved, and only excitatory activity is used for reconstruction, and quantify the influence of the inhibitory population on the estimation (Figure 4C). This approach allows us to examine the impact of inhibitory dynamics on effective heterogeneity and their role in shaping network-level properties.

We first characterize the reconstruction method of an excitatory-inhibitory model (Model C, Eqs. 37-42) as similar in section II [38,65].

The linearized dynamics around the steady states $S^*$ [Materials and Method]:

$$\frac{dS}{dt} = J_{S^*}(S - S^*) + \sigma v(t),$$ (10)

where $S = \begin{bmatrix} S_E \\ S_I \end{bmatrix}_{2N \times 1}$, and $J_{S^*}$ is the full Jacobian matrix in the form of a block matrix [Materials and Method].

Here, $N = 29$ represents the number of cortical areas. For each cortical region with two populations (E/I), the Jacobian matrix $J_{S^*}$ has $2N$ dimensions with four $N$-dimension blocks: $\begin{bmatrix} J_{EE} & J_{EI} \\ J_{IE} & J_{II} \end{bmatrix}_{2N \times 2N}$ where $J_{pq}$ denotes the connections from population $q$ to $p$, where $p, q \in \{E, I\}$. The top-left block $J_{EE}$ represents the long-range excitation-to-excitation connectivity across regions, while $J_{EI}, J_{IE}$ and $J_{II}$ are diagonal matrices since only excitatory populations participate in inter-regional communications (Eq.48).

We then derive the approximated ground truth $J_{eff}$ of excitatory population $S_E$ [Materials and Method]:

$$J_{eff} = J_{EE} - J_{EI} J_{II}^{-1} J_{IE}. \tag{11}$$

Because $J_{EI}, J_{IE}$ and $J_{II}$ are diagonal matrices (there is only long-range E-E connection), this absorption indicates that localized inhibitory populations mainly modulate the parameters of corresponding excitatory population of each region. Following Eq. 11, the ground truth effective Jacobian $J_{eff}$ can be represented as effective heterogeneity and asymmetric SC (same as Eq. 5):

$$\left[ J_{eff} \right]_{ij} = h_i C_{ij}, \tag{12}$$

with effective heterogeneity

$$h_i = \gamma G \left( 1 - S_E^{i*} \right) g_E^i, \tag{13}$$

where $g_E^i = \frac{dH_E^i}{dx_E^i} \big|_{x_E^{i*}}$ is the firing rate change of excitatory population (gain function level).

Overall, Eq. 11 provides an approximated mathematical ground truth during the reconstruction when using only the excitatory activity $S_E$, enabling us to further test the separation of effective heterogeneity and asymmetric structural connectivity (Eqs. 7-9, Figure 4D).

Previous studies have found that the best fit to empirical data occurs at the edge of Hopf bifurcation, representing a regime where stable dynamics coexist with oscillatory behavior [64,65]. To validate our prediction on effective Jacobian in Eq. 11 and whether our reconstruction method can still hold with only excitatory activity observations and symmetric SC, we simulated the model C with or without the feedback inhibition control (FIC, controlling that regional firing rates are ~3Hz, see also Eq. S4) and only kept the excitatory activity $S_E^i$ of each region for the temporal (Eq. 7) and spatial reconstruction (Eq. 9).

Simulation results indicated good reconstruction performance in estimating the off-diagonal elements of effective Jacobian $J_{eff}$ using only excitatory activity (Figure 4D and Figure S6). Also, the diagonal elements of the estimated effective Jacobian are well aligned with our approximation in Eq. 11, further validating the precision of our approximation (Figure S7).

Similarly, we characterized bifurcations in Model C under both FIC and non-FIC conditions. Our results demonstrate that even at bifurcations, we can achieve sufficiently good effective heterogeneity and asymmetry reconstruction (Figure S6). Although we noted that determining the effective heterogeneity $h_i$ and asymmetric SC $C_{ij}$ only requires the excitatory activity $S_E^i$ and symmetric SC $W_{ij}$, achieving further detailed reconstruction of parameterized heterogeneity $w_{EE}^i$ and $w_{IE}^i$ remains a challenge due to the lack of inhibitory population parameters.

Together, our results highlight the feasibility of mapping different parameterized heterogeneities onto their functional contributions as distinct components of effective heterogeneity, and demonstrate that this reconstruction method performs well in E-I networks with hidden local inhibitory populations. These analyses support the flexibility of heterogeneous modeling approaches in capturing the functional dynamics of large-scale brain networks.

## VI Sampling Interval Effect on Estimating Effective Connectivity

A key question is how the underlying neural interactions inferred from neural activity depend on the temporal resolution of empirical data, which is hardly noticed in EC studies [26,44]. This mismatch between temporal resolutions might bring system errors in multi-modality comparisons (like MEG and fMRI signals) and pose significant challenges to the interpretation of the obtained EC. To address this, we investigated the influence of sampling resolution of the neural activity on temporal reconstruction performance and mathematically quantified the relationship between the sampling rate and the reconstructed Jacobian matrix. We began by discretizing the linear system described in Eq.4 with a time step $h$:

$$\frac{x(t+h) - x(t)}{h} = J_o x(t) + \sigma v(t),  \tag{14}$$

where $x(t) = S(t) - S^*$ represents the shifted neural activity and $J_o = J$ is the original Jacobian matrix at the stable fixed point $S^*$.

Due to technical limitations (e.g., fMRI detects BOLD signals at intervals on the order of hundreds of milliseconds), observed neural activity is sampled at a sampling step $n$ with a time step $T$. Under these conditions, the discrete linear system can be represented in an exponentially diffused manner [Materials and Method] [44]:

$$\frac{x(t+T) - x(t)}{T} \approx \frac{e^{TJ_o} - I}{T} x(t) + B_T v(t). \qquad (15)$$

The ground truth Jacobian matrix after sampling is expressed as:

$$J_T = \frac{e^{TJ_o} - I}{T}, \qquad (16)$$

and the matrix estimated by temporal reconstruction is described as:

$$\hat{J}_T = <\frac{x(t+T) - x(t)}{T}, x(t) > < x(t), x(t) >^{-1}. \qquad (17)$$

Eq.16 illustrates how the temporal resolution of the sampled activity results in an exponential scaling of the true Jacobian matrix $J_o$ at a rate determined by the sampled time step $T$. This occurs because the true dynamics evolve in the network between one sampling time step and the next, a phenomenon previously noted in studies inferring EC from fMRI time series [44]. Therefore, we have provided a ground truth EC under the sampling interval $T$, while using sampled neural activity for temporal reconstruction.

Then, the estimation of the true Jacobian matrix $\hat{J}_o$ is solved from Eq. 17 using matrix logarithm operation [Materials and Method]:

$$\hat{J}_o = \frac{\ln\left(T\hat{J}_T + I\right)}{T}. \tag{18}$$

We assessed the effect of sampling interval by examining the reconstructed matrix $\hat{J}_T$ from sampled neural activity across sampling time steps $T$ ranging from 0.01s to 1s. The estimated matrix $\hat{J}_T$ was strongly correlated with the deduced exponential diffusion matrix $J_T$ in Eq. 16 across different sampling steps (Figure 5A).

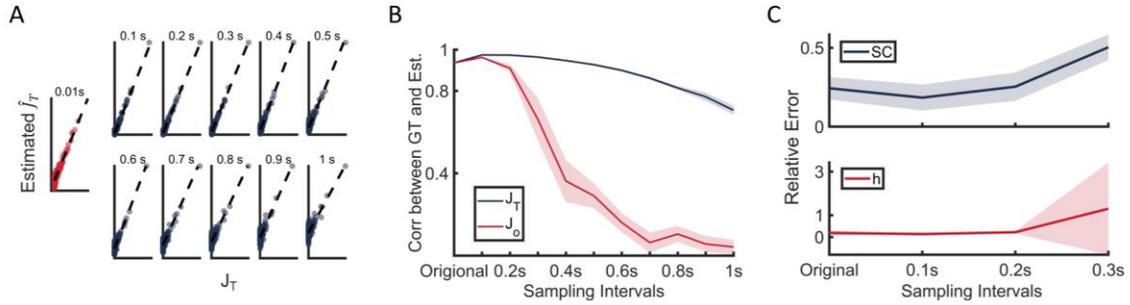

**Figure 5. Reconstruction performance with respect to the sampling interval of observations.** (A and B) Reconstruction and prediction of exponential scaling strongly match across sampling steps. (A) Element-wise comparison between estimated Exponential Jacobian $\hat{J}_T$ from sampled neural activity and analytical Jacobian $J_T$ across different sampling resolution from 0.01s (unsampled, red) to 1s (100 steps, blue). Dash line represents $y = x$. (B) Correlation of $\hat{J}_T$ and $J_T$ (blue), $\hat{J}_o$ and $J_o$ (red) across different sampling step $n$. (C) The relative errors between ground truth and estimation of asymmetric SC (blue) and effective heterogeneity $h_i$ (red) estimated from $\hat{J}_o$ as functions of sampling steps. The colored lines show the mean relative errors across 10 simulations, with shaded areas indicating one standard deviation from the mean. Simulation length $50,000s$.

Both correlations of $\hat{J}_T$ and $J_T$, $\hat{J}_o$ and $J_o$ decreased as the sampling interval increased (Figure 5B). Notably, the correlation between the ground truth $J_o$ and the estimated $\hat{J}_o$ (Figure 5B, red) declines more rapidly than that between $J_T$ and $\hat{J}_T$ (Figure 5B, blue) at sampling interval of 0.3s or longer. This stronger decline can be attributed to two key factors. First, downsampling preferentially preserves strong connections while weak connections decay toward zero, as temporal precision and rapid neural activity changes are lost in the process. This effect, combined with estimation noise, leads to increased false positive rates at longer sampling intervals (Figure 5A, evidenced by greater scatter near zero). The transformation from $\hat{J}_T$ to $\hat{J}_o$ through the matrix logarithm operation further amplifies these effects (Figure 5B, red), reflecting the fundamental limitation in reconstructing network connectivity from low-temporal-resolution observations. Second, the estimation $\hat{J}_o = \frac{\ln(T\hat{J}_T + I)}{T} \in C^{N \times N}$ is not uniquely defined under the matrix logarithm [44]. The solution's uniqueness is determined by both the asymmetry of $\hat{J}_T$ and the sampling interval $T$, leading to increased sparsity in $J_T$ and consequently, non-unique solutions.

Further reconstruction of effective heterogeneity $h_i$ and asymmetric SC $c_{ij}$ from $\hat{J}_o$ demonstrated high precision in estimation for up to sampling interval of 0.2s (Figure 5C). However, the precision and stability of these estimations deteriorated at 0.3s, likely due to reduced accuracy in estimating the true Jacobian matrix $\hat{J}_o$.

# Discussion:

Local heterogeneity and asymmetry connections jointly shape the directionality of information flow in dynamical neural networks. In this study, we developed a reconstruction framework based on the existing directionality estimation method DDC to further separate local heterogeneity from asymmetric connections. Evaluation on dynamical mean field models as ground truth has shown that our approach effectively recognizes the existence of local heterogeneity and asymmetric connections through the directionality of the estimated Jacobian matrix and symmetric connection constraints. The separated heterogeneity revealed by this method quantifies how different forms of parameterized local heterogeneity together with the asymmetric connections alter whole brain dynamics. We also deduced a theoretical prediction on the effective interactions between excitatory populations with hidden inhibitory populations. Finally, we demonstrated the sampling interval effect with respect to temporal resolution in sampling the neural activity in observations in a linear dynamic framework. Collectively, our approach highlights the potential for segregating biophysical asymmetric structural connections and regional heterogeneity from neural activity.

## The Entangled Local Heterogeneity and Asymmetric Connections

Regional heterogeneity is crucial for generating and guiding information flow within the cortex. It supports functional segregation across structurally separated brain regions [37,38,43,50]. Experimental evidence suggests that regions with higher average pyramidal neuronal spine counts are thought to exhibit greater self-excitation and higher probability to act as source of activation spreading, representing the directionality of information transmission, enhancing information flow and contributing to the directionality of functional connectome [8,43,51]. This information directionality is also based on asymmetric SC, which captures richer feedforward and feedback details at the inter-regional level. This suggests that regional heterogeneity and asymmetric anatomical connections are entangled together, but little has been discussed in how they determine the directionality of the functional patterns [16-19]. Here, we propose a method that can separate and simultaneously estimate regional heterogeneity and asymmetrical structural connections. Specifically, this method estimates the directionality within the connectivity matrix and subsequently strips the regional heterogeneity of the brain network. Previously, based on the assumption of uniform regional dynamics, several theoretical approaches have been developed to estimate the directionality of connectomes, ranging from simple statistical inferring from functional connectivity to more sophisticated dynamical causal modelling leveraging hidden information within neural activity to uncover the directionality underlying neural activity [25-35].

It is important to note that our framework performs optimally in linear regimes where nonlinear interactions are not too strong. In highly nonlinear regimes, particularly where complex bifurcation dynamics emerge, parameter estimation becomes more challenging due to the nonlinear regimes in firing rate functions. However, even under these conditions, our approach maintains reasonable performance in capturing functional connectivity patterns, while still providing mechanistic insights through the decomposition of EC into effective heterogeneity and asymmetric SC.

Our method thus highlights the necessity to consider both regional heterogeneity and asymmetrical connections, thereby can be used for further probing their distinct functional roles in the whole brain dynamics.

**Mapping Anatomical and Functional Heterogeneity**

Separated regional effective heterogeneity provides a quantitative mapping between anatomical and functional heterogeneity. While imaging studies have empirically detected various forms of anatomical heterogeneity [1-9,20] and are increasingly incorporating this data into large-scale circuit models [36-39,43,50], these studies have not been considered within a unified dynamical framework. Specifically, Chaudhuri et al. organized cortical areas in macaques based on laminar feedforward and feedback connections, embedding this heterogeneity into the excitatory input for all areas [43,50]. In contrast, Demirtaş et al. incorporated heterogeneity by embedding the T1w/T2w ratio into local recurrent strength [38], while Deco et al. focused on the role of gene expression of excitatory and inhibitory receptors in modifying the curvature of the activation function [37]. It was also shown that the gradient of synaptic inhibition, not the gradient of recurrent excitation, can well characterize the persistent activity patterns in the mouse cortex [39]. This raises the question: how different anatomical heterogeneity of intrinsic cortical features shape whole-brain dynamics?

We proposed to shift the focus to whether there are constraint rules for defining heterogeneous parameters in shaping dynamics. Our analysis revealed that effective heterogeneity can be systematically separated across different parameterized heterogeneity patterns. This consistency suggests a fundamental cause-and-effect relationship between anatomical and functional hierarchies, providing a framework for understanding mapping between different heterogeneity configurations. Importantly, we found that effective heterogeneity can be modulated by hidden inhibitory neuronal populations, which influence the overall level of network self-recurrence. This is equivalent to downscaling the entire network, while pairwise correlations are preserved through adjustments in the effective heterogeneity of EC [60]. Our method provides insights into how different parameterizations of heterogeneity can lead to similar functional outcomes in neural circuits.

While our linearized framework operates primarily in linear regimes, it successfully decomposes directional connectivity into neural activity-dependent effective heterogeneity and activity-independent asymmetric structural connectivity. Crucially, even in highly nonlinear regimes where parameter estimation becomes challenging, our approach maintains the ability to capture essential functional connectivity patterns, demonstrating robustness across different dynamical states. Although feedback inhibition control and the unique role of E-I populations (such as balanced amplification and oscillatory dynamics) are critical features that cannot be fully captured by linear approximations, our framework still provides a valuable approach for separating and estimating effective heterogeneity and asymmetric structural connectivity, offering fundamental insights into the constraint rules governing heterogeneous parameter organization and their impact on network dynamics.

## Challenges in Estimating Directionality: Sampling Interval Effects of sampling rate and Hemodynamics

Temporal resolution in sampling inconsistency of neural activity resolution leads to estimation bias. fMRI data from human subjects have facilitated the application of established theoretical methods to estimate the directionality of brain networks [26,27,33-35,52]. While this approach is prevalent in imaging neuroscience, BOLD signals suffer from poor temporal resolution, approximately 0.7-2 seconds [24-26]. This limitation can hinder the accurate detection of rapid neural dynamics and may compromise the reliability of directionality estimates, leading to spurious connections and false positives. Moreover, previous network inference studies [32,33,44,53] rarely consider the effects of sampling rate on the performance of estimating ground truth connectivity, which has limited their application to real data.

Here, we identified a sampling interval effect: the EC estimated from sampled data exhibits an exponential scaling relationship with the true SC. This effect occurs when there is a temporal resolution mismatch between the real neural activity (which we assume has dynamics with high temporal resolution) and the observed neural activity (sampled from the real neural activity at given intervals). For instance, when contrasting fMRI with MEG, the discrepancies may include sampling interval effects, thereby reducing the accuracy of EC estimation. This creates two challenges: (1) exponential scaling transforms the original structural connectivity into sampling interval dependent EC, causing weaker connections to decay while it preserves only the strongest interactions, and (2) the matrix logarithm transformation amplifies noise and can yield non-unique solutions to recover the original structural connectivity at longer intervals. Recent work has found that whole-brain pulse-response probability can be largely estimated by similar exponential scaling forms of SC, suggesting another interpretation of our work: the sampling interval effect also reveals the manner of information transmission during temporal intervals [66].

Despite active discussions in the engineering literature on overcoming this sampling issue—such as addressing the Nyquist frequency of the sampling rate and natural fluctuation rates—key aspects, such as comparing estimation methods with or without the matrix logarithm, remain underexplored and deserve further investigation [44]. Additionally, the estimation of matrix logarithm introduces technical challenges, as the existence and uniqueness of valid solutions are highly dependent on the structure of the connectivity matrix. Future directions may focus on developing approximation strategies and regularization techniques for matrix logarithm computation, which could potentially yield more accurate estimators and ensure robust solutions across different connectivity matrix structures. Consequently, although fMRI provides valuable insights into brain connectivity, the effects of BOLD signal resolution require careful consideration when interpreting findings related to brain network directionality.

In addition to the sampling interval effect, BOLD fMRI data presents another challenge: the hemodynamic response function convolves the underlying neural activity, creating temporal dependencies that obscure the direct relationship between neural dynamics and observed BOLD signals [54-56]. This hemodynamic convolution process systematically biases effective connectivity estimation because the observed BOLD signals represent a temporally smoothed and delayed version of the actual neural activity. The convolution process complicates the recovery of neural connectivity patterns, particularly for asymmetric connections where directionality and timing are crucial [44]. To address this limitation, future studies should incorporate hemodynamic response modeling directly into the connectivity estimation framework. This could be achieved through either deconvolution approaches that attempt to recover neural signals from BOLD data, or forward modeling approaches that explicitly account for hemodynamic convolution effects during connectivity estimation. Such methods have proven successful in calcium imaging, where deconvolution techniques are routinely used to infer neural activity from fluorescence signals [57-59].

**Conclusion and Outlooks**

In conclusion, we proposed a method to simultaneously estimate local heterogeneity and asymmetric connections from observation of neural dynamics. Our findings provide a theoretical framework for further analysis of relative contributions of local heterogeneity and asymmetric connections to network dynamics.

Looking ahead, we will integrate multi-region MEG/sEEG datasets having high temporal resolution with symmetric structural connectivity from dMRI to uncover directional interactions between cortical areas and elucidate how local heterogeneity influences global brain function. We will pursue empirical validation by testing whether our effective heterogeneity measure correlates with region-specific features such as intrinsic timescales and neural variability in spiking dataset [69,70]. We suggest that effective heterogeneity may reflect the directionality of neural information flow, including top-down and bottom-up pathways, with directional flow switching potentially captured through variance in regional effective heterogeneity [71-73]. Additionally, we aim to investigate how the obtained asymmetric SC is related to the modular organization of the cortex – a key determinant of its functional segregation and integration [61,62].

By integrating our reconstruction framework with multimodal neuroimaging data, we can bridge gaps between anatomical organization and functional dynamics, ultimately advancing mechanistic models of how neural circuitry supports cognition and behavior.

# Materials and Method

## 1. Anatomical Data

**1.A Anatomical Connectivity.** The asymmetric SC used in dynamical modeling of the brain network is derived from a comprehensive project to quantitatively characterize all inter-areal connections in the macaque cortex [10], utilizing retrograde tracer injections to label projecting neurons and measuring connection strengths as fractional weights (fraction of labeled neurons). The anatomical connection from area $j$ to area $i$ is defined as the number of neurons projecting from area $j$ to area $i$, normalized by the total number of neurons projecting from all areas to area $i$:

$$C_{ij} = \frac{\#\ neurons\ projecting\ from\ area\ j\ to\ area\ i}{\#\ neurons\ projecting\ from\ all\ areas\ to\ area\ i}.$$

In this research, we use 29 cortical areas reported from previous large-scale modeling studies that explored how asymmetric SC and anatomical hierarchy shape large-scale cortical dynamics [43,50]. The symmetric SC is defined as the average of asymmetric SC: $W = \frac{C+C^T}{2}$. This assumption aligns with dMRI-based SC, which can only detect fiber existence and density [21-23].

The asymmetry level $\eta(C)$ of the connectivity matrix $C$ is calculated as the element-wise correlation between the upper and lower triangular matrices, quantifying the linear relationship between feedforward and feedback connections [63]: $\eta(C) = corr(C_{i>j}, C_{j>i})$.

We report that the asymmetry level of empirical macaque connectivity $\eta(C)$ is 0.7018, which represents that the connectivity matrix is symmetry-dominated but regulated by asymmetric connections.

**1.B Regional Heterogeneity Implementation.** The anatomical hierarchy is derived from the same dataset [15] using a generalized linear model that assigns hierarchical values based on the supragranular layer neuron fraction between cortical areas [15,43]. This anatomical hierarchy strongly correlates with T1w/T2w maps reflecting myelination [8]. Considering that T1w/T2w mapping is widely utilized in heterogeneous large-scale cortical modeling studies [36-38], we parameterized this anatomical hierarchy across the macaque cortex by incorporating region-specific local recurrent strength $w_i$ into the model. External inputs $I_i$ are chosen to sequentially decrease along the anatomical hierarchy, which are suggested to reflect the flow of sensory information from the external environment [36].

The range of $w_i$ and $I_i$ are scaled to $[0.0652,0.1581]nA$ and $[0.33,0.3]nA$ according to previous study on the same model [36].

## 2. Mathematical Analysis

### 2.A Detailed Jacobian elements of Model A.

In this section, we derive the Jacobian matrix of Model A (Eqs. 1-3) around its steady state through Taylor expansion, retaining only the first-order terms. We further elaborate on the biological interpretation of the Jacobian matrix.

We start with the noise-driven nonlinear Model A (Eqs. 1-3) and write it into a general nonlinear dynamic:

$$\frac{dS}{dt} = f(S; \theta) + \sigma \nu(t),$$

where $\theta$ represents system parameter and $\nu(t)$ is $i.i.d.$ white noise. We then use Taylor expansion around steady state $S^*$ and keep the first-order terms:

$$\frac{dS}{dt} = f(S^*; \theta) + \frac{\partial f(S; \theta)}{\partial S}|_{S=S^*}(S - S^*) + o\left(\frac{\partial f(S; \theta)}{\partial S}\right) + \sigma \nu(t)$$

$$\approx \frac{\partial f(S; \theta)}{\partial S}|_{S=S^*}(S - S^*) + \sigma \nu(t)$$

$$= J(S^*)(S - S^*) + \sigma \nu(t).$$

Here in the first line, $o(\frac{\partial f(S; \theta)}{\partial S})$ is the higher-order terms of Taylor expansion in the form of Peano's Remainder and is neglected in the following derivation. The Jacobian matrix $J(S^*) = \frac{\partial f(S; \theta)}{\partial S}|_{S=S^*}$ is the first order derivative of nonlinear function $f$ near steady state $S^*$. The elements of Jacobian matrix are:

$$\begin{cases} J_{ii}(S^*) = -\dfrac{1}{\tau_S} - \gamma H(x_i)|_{x_i=x_i^*} + w_i \gamma (1 - S_i^*) \dfrac{\partial H}{\partial x_i}|_{x_i=x_i^*}, i = j \\ \qquad J_{ij}(S^*) = \gamma G(1 - S_i^*) C_{ij} \dfrac{\partial H}{\partial x_i}|_{x_i=x_i^*}, i \neq j \end{cases} . \qquad (19)$$

The diagonal elements ($i = j$) describe how regional heterogeneity affects the self-regulation of each region: inherent time constant $\tau_S$ and regional firing rate $H(x_i)$, and the last term indicating recurrent excitation $w_i$, regional stable states $S_i^*$ and regional firing rate change $\frac{dH}{dx}|_{x_i}$ entangle together to regulate the regional decay rate. Also, recurrent excitation $w_i$ and external input $I_i$ both contribute to the regional overall input $x_i^*$, therefore affect the regional firing rate $H(x_i)$ and firing rate change $\frac{dH}{dx}\big|_{x_i = x_i^*}$.

The off-diagonal elements $(i \neq j)$ describe that the long-range communication from region $j$ to region $i$ is regulated by both the asymmetric projections $C_{ij}$ and stable state $\{S_i^*, x_i^*\}$ of target region $i$. We noted that the difference of regional stable states still exists even in the absence of heterogeneity and asymmetry; this is because the graph properties (e.g., degree) of each region are different, but asymmetry and heterogeneity further enrich these dynamics [36-41,43].

## 2.B Procedure of Temporal Reconstruction and Spatial Reconstruction

Here, we provide mathematical details of how we process temporal reconstruction using DDC in Eq. 7 and spatial reconstruction to separate EC into effective heterogeneity and asymmetric SC in Eqs. We first took the outer product on both sides of Eq. 4 with $S$ and average over time, yielding:

$$< \frac{dS}{dt}, S >= J < S, S >, \tag{20}$$

where $J$ represents $J(S^*)$, $< S, S > = < S - S^*, S >$ is the covariance matrix of neural activity, and $< v, S > \approx 0$ and is ignored, as shown in reference [35]. Here, the angle brackets $<, >$ denote time averages.

Taking the inverse of covariance $< S, S >^{-1}$ on both sides, we derive the estimator for the Jacobian matrix which is asymmetrical in Eq. 7.

We next derive the multivariate least squares problem of spatial reconstruction in Eqs. 8 and 9. We first reform Eq. 8 with $y_i = \frac{1}{h_i}$:

$$\hat{J}_{ij} y_i + \hat{J}_{ji} y_j = 2W_{ij},$$

and the corresponding vector form

$$M y = W_{vec}, \tag{21}$$

where $y_{N \times 1} = [y_1, y_2, \ldots, y_N]^T$, $M \in \mathcal{R}^{(N^2-N) \times N}$ is reform of $\hat{J}_{ij}$ and $\hat{J}_{ji}$, and $W_{vec} \in \mathcal{R}^{(N^2-N) \times 1}$ is the vectorized form of the non−diagonal elements of $W$. The estimation of $y$ can then be solved using the following closed-form solution:

$$\hat{y} = (M^T M)^{-1} M^T W_{vec}. \tag{22}$$

Using $\hat{y}$, we can proceed to estimate effective heterogeneity $\hat{h}_i = \frac{1}{\hat{y}_i}$, and $\hat{C}_{ij} = \hat{y}_i \hat{J}_{ij}$.

## 2.C Spatial Reconstruction Without Effective Heterogeneity

To provide a baseline comparison for our method of simultaneously separating effective heterogeneity and asymmetric SC from estimated EC (i.e., Spatial Reconstruction), we first derived an optimization approach that ignores effective heterogeneity and considers only asymmetric SC. In this case, the asymmetry of Jacobian matrix is solely contributed from the asymmetry of SC and scaled by a constant $h$. In this baseline method, the inputs are the estimated EC (Jacobian) and symmetric SC, while the output is only the asymmetric SC.

We adjusted our original spatial reconstruction (Eqs. 8 and 9) by treating the effective heterogeneity $h_i$ to be homogeneous across regions:

$$\begin{cases} hC_{ij} = \hat{J}_{ij} \\ \dfrac{C_{ij} + C_{ji}}{2} = W_{ij} \end{cases}, i \neq j. \tag{23}$$

We then write the loss function $L$ for optimizing $C$:

$$L(C; J, W) = \|J - hC\|_F^2 + \|C + C^T - 2W\|_F^2, \tag{24}$$

where $\|\cdot\|_F^2$ is the square of Frobenius norm quantifying the sum of least squares errors of matrix elements. The optimization was done by using *fmincon.m* in MATLAB.

The relative errors $RE(C_{ij}, \hat{C}_{ij})$ are calculated across each $G$ value for benchmark comparison during reconstruction of asymmetric SC.

## 2.D Spatial Reconstruction Without Asymmetry

Above, we derived a baseline comparison that systematically ignores effective heterogeneity. Here, we consider an alternative baseline method where the directionality is entirely contributed by effective heterogeneity, while assuming the ground truth SC is symmetric. We adjusted the spatial reconstruction method (Eqs. 8 and 9) by treating the asymmetric SC $C_{ij}$ to be symmetric:

$$h_i W_{ij} = \hat{J}_{ij}, i \neq j. \tag{25}$$

In this case, we assume that the asymmetry of Jacobian matrix is purely contributed from the regional effective heterogeneity $h_i$. The estimation of each $h_i$ is calculated:

$$\hat{h}_i = \left(W_{i,:}^T W_{i,:}\right)^{-1} W_{i,:}^T \hat{J}_{i,:},$$

where $W_{i,:} \in \mathcal{R}^{1,N}$ is the $i$-th row of symmetric SC $W_{ij}$, $\hat{J}_{i,:} \in \mathcal{R}^{1,N}$ is the $i$-th row of Jacobian $\hat{J}_{ij}$.

The relative errors between $RE(\hat{h}_i, h_i)$ are calculated for benchmark comparison during reconstruction of effective heterogeneity.

## 2.E Heterogeneity and Asymmetry Coupling Reconstruction - Detailed Spatial Reconstruction of Model A.

To further reconstruct the local circuit heterogeneity $w_i$, $I_i$ and directed SC $C_{ij}$ based on the estimated Jacobian matrix (EC) in the form of model A, we first need to approximate the diagonal terms of Jacobian matrix. By assuming $\frac{dS_i^*}{dt} \approx 0$ and neglecting the effect of noise input $\sigma v_i(t)$ in Eq.1, we have:

$$\gamma H(x_i^*) = \frac{S_i^*}{\tau_s(1 - S_i^*)} \tag{26}$$

Taking Eq. 26 into Eq.5 and Eq.6, the Jacobian Matrix $J(S^*)$ then can be expressed as:

$$\begin{cases} J_{ii}(S^*) = -\dfrac{1}{\tau_S(1 - S_i^*)} + \dfrac{w_i h_i}{G}, i = j \\ J_{ij}(S^*) = h_i C_{ij}, i \neq j \end{cases} . \tag{27}$$

This approximation holds since the effect of noise fluctuation is relatively small near the fixed point $S^*$.

Therefore, with Eq.26 and Eq.27, and recall that $y_i = \frac{1}{h_i} = \frac{1}{\gamma G J_N (1 - S_i^*) \frac{\partial H}{\partial x_i}|_{x_i = x_i^*}}$, we can solve the local recurrent strength:

$$\widehat{w}_i = G \hat{y}_i \hat{J}_{ii} + \frac{G \hat{y}_i}{\tau_S(1 - S_i^*)}. \tag{28}$$

To reveal regional external input $I_i$, we use *fsolve.m* in MATLAB to solve for $x_i^*$ from the derivative of activation function:

$$\frac{\partial H}{\partial x_i}\Big|_{x_i = x_i^*} - \frac{1}{\gamma G (1 - S_i^*) \hat{y}_i} = 0. \tag{29}$$

After estimating $x_i^*$ and using Eq.3, the regional external input $I_i$ is calculated by the following formula:

$$\hat{I}_i = x_i^* - \hat{w}_i S_i^* - G\Sigma_j \hat{C}_{ij} S_j^*. \tag{30}$$

## 2.F Detailed Spatial Reconstruction of Model B.

We seek to probe the ability of detailed reconstruction in an alternative large-scale circuit model (Model B) adapted from *Deco et al* [37]. From the perspective that the ability of accumulating and leaking information of each region varies, we define timescale $\tau_i$ and firing threshold $b_i$ are heterogeneous in this model while keeping the local recurrent strength $w$ and external input $I$ homogeneous:

$$\frac{dS_i(t)}{dt} = -\frac{S_i}{\tau_i} + \gamma(1 - S_i)H(x_i) + \sigma v_i(t) \tag{31}$$

$$H_i(x_i) = \frac{ax_i - b_i}{1 - \exp(-d(ax_i - b_i))} \tag{32}$$

$$x_i = wS_i + G\Sigma_j C_{ij} S_j + I, \tag{33}$$

where $\tau_i$ and $b_i$ can represent the ability of leaking and storing information of each ROI.

Similarly, the Jacobian matrix $J_B$ of Model B near steady state $S^*$ in Eqs. 31-33 can be derived:

$$\begin{cases} J_{B,ii}(S^*) = -\dfrac{1}{\tau_i(1-S_i^*)} + w\gamma(1-S_i^*)\dfrac{\partial H_i}{\partial x_i}\big|_{x_i=x_i'}, i = j \\ \qquad J_{B,ij}(S^*) = \gamma G(1-S_i^*)C_{ij}\dfrac{\partial H_i}{\partial x_i}\big|_{x_i=x_i'}, i \neq j \end{cases} \quad . \tag{34}$$

We noted that the diagonal terms of $J_B$ are approximated by assuming $\frac{dS_i^*}{dt} \approx 0$. In Model B, activation function $H_i$ is heterogeneous across ROIs because of $b_i$ and we defined $x_i' = wS_i^* + G\Sigma_j C_{ij}S_j^* + I$.

The detailed reconstruction follows the general reconstruction procedure of temporal reconstruction and spatial reconstruction as shown in Eq. 9 to reveal effective heterogeneity $h_i^B$ and asymmetric connections $\hat{C}_{ij}$, we can therefore estimate

$$\frac{1}{\hat{\tau}_i} = (1-S_i^*)\left(\frac{wh_i^B}{G} - J_{B,ii}\right), \tag{35}$$

and we also use *fsolve.m* in MATLAB to solve $b_i$ from the derivative of activation function:

$$\frac{\partial H_i}{\partial x_i}\big|_{x_i=x_i'} - \frac{h_i^B}{\gamma G(1-S_i^*)} = 0. \tag{36}$$

The ground truth $\tau_i$ and $b_i$ of Model B are calculated from Eq. S2 using ground truth $w_i$ and $I_i$ of Model A, which provides a mapping relationship between these parameter pairs. The homogeneous $w$ and $I$ of Model B are the average of ground truth $w_i$ and $I_i$ of Model A.

## 2.G Absorption of Local Inhibitory Populations into Effective Jacobian - Model C.

Fluctuations in whole-brain imaging data such as BOLD and MEG signals are thought to largely reflect excitatory activity rather than inhibitory activity [44,48]. When estimating EC using such empirical data, the role of inhibitory populations is systematically neglected. A fundamental question arises: what exactly is the EC being reconstructed under these unobserved inhibitory dynamics? In this section, using Model C (E-I model) [38], we first derive an approximate solution for the ground truth EC when only excitatory activity is considered, and demonstrate how the estimated EC is affected by E-I interactions and subsequent decomposition for effective heterogeneity and asymmetric SC.

We introduce Model C considering both excitatory and inhibitory populations:

$$\frac{dS_E^i(t)}{dt} = -\frac{S_E^i}{\tau_E} + \gamma\left(1 - S_E^i\right)H_E\left(x_E^i\right) + \sigma v_E^i(t) \tag{37}$$

$$\frac{dS_I^i(t)}{dt} = -\frac{S_I^i}{\tau_I} + H_I\left(x_I^i\right) + \sigma v_I^i(t) \tag{38}$$

$$H_E\left(x_E^i\right) = \frac{a_E x_E^i - b_E}{1 - \exp\left(-d_E\left(a_E x_E^i - b_E\right)\right)} \tag{39}$$

$$H_I\left(x_I^i\right) = \frac{a_I x_I^i - b_I}{1 - \exp\left(-d_I\left(a_I x_I^i - b_I\right)\right)} \tag{40}$$

$$x_E^i = w_{EE}^i S_E^i + G\Sigma_j C_{ij} S_E^j - w_{EI} S_I^i + I_E, \tag{41}$$

$$x_I^i = w_{IE}^i S_E^i - S_I^i + I_I. \tag{42}$$

All parameters have the same properties as in model A and B but different in their values as listed in Table 3. For this model, we assume that the inhibition connections $w_{EI}$, are known and homogenous for different ROIs, and the local heterogeneity is attributed to excitatory connection $w_{EE}^i$ and $w_{IE}^i$ [38]. Here, we set $w_{EE}^i$ and $w_{IE}^i$ to increase linearly along the same anatomical hierarchy (Figure 3A) (Table 3).

**Table 3 Fixed Parameters for Large-Scale Excitatory-Inhibitory Circuit Model (Model C)**

| Parameter | Excitatory Populations | Inhibitory Populations |
| --- | --- | --- |
| $\tau_P$ | $0.1\ s$ | $0.01\ s$ |
| $a_P$ | $310\ nC^{-1}$ | $615\ nC^{-1}$ |
| $b_P$ | $125\ Hz$ | $177\ Hz$ |
| $d_P$ | $0.16\ s$ | $0.087\ s$ |
| $I_P$ | $0.382\ nA$ | $0.2674\ nA$ |
| $w_{EI}$ | 1 | - |
| $\gamma$ | 0.641 | - |
| $w_{PE}^i$ | $0.126-0.210\ nA$ | $0.090-0.150\ nA$ |

To characterize the EC of the excitatory population in the presence of unobserved inhibitory populations, we mathematically reformulate the problem as an effective linear dynamical description of the excitatory population:

$$\frac{dS_E}{dt} = J_{eff}(S_E - S^*{}_E) + \sigma\nu_E(t), \tag{43}$$

where $J_{eff}$ is the effective Jacobian matrix neglecting the modulation from inhibitory population as comared with the whole E-I network (Eqs. 37-42). Consequently, when estimating EC using only excitatory activity, we can obtain an unbiased estimate of $J_{eff}$. Next, we seek to find the relationship between the ground truth effective Jacobian $J_{eff}$ and the full Jacobian $J_{S^*}$, then the effective heterogeneity $h_i$ and asymmetric SC $C_{ij}$ under $J_{eff}$, as we did during the spatial reconstruction. Crucially, while $J_{eff}$ and $J_{S^*}$ both capture the same pairwise correlations within the excitatory population, they differ in dimensionality. We therefore derive the ground truth $J_{eff}$ by leveraging the relationship between the Jacobian matrix and the covariance structure (unnormalized FC), which serves as the ground truth benchmark for evaluating our reconstruction approach in Section V.

According to Eq.20, the linearized dynamics of $S_E$ can be expressed as:

$$\frac{dS_E}{dt} = J_{EE}S_E + J_{EI}S_I + \sigma\nu_E,$$

take outer product of $S_E$ and average over time, we have

$$< \frac{dS_E}{dt}, S_E > = J_{EE} < S_E, S_E > + J_{EI} < S_I, S_E >,$$

then take the inverse of $COV_{EE} = < S_E, S_E >$, we can express $J_{eff}$ as:

$$J_{eff} = < \frac{dS_E}{dt}, S_E > < S_E, S_E >^{-1} = J_{EE} + J_{EI}COV_{IE}COV_{EE}^{-1}. \tag{44}$$

Compared to the estimation of reduced dynamics (Model A, Eq.9), the estimation in Eq. 45 indicates that the inhibitory population in the dynamics will introduce a bias term in estimating EC: $J_{EI} COV_{IE} COV_{EE}^{-1}$. To illustrate what underlying matrix the $J_{eff}$ is representing, we need to represent the covariance $COV_{IE}$ and $COV_{EE}^{-1}$ with block matrices of the full Jacobian $J_{S^*}$.

To do so, we can represent the block matrices of the full covariance matrix $\Sigma$ with block matrices of Jacobian $J_{S^*}$ by approximating $\Sigma = \sigma^2 J_{S^*}^{-1} J_{S^*}^{-T}$: letting $\frac{dS}{dt} = 0$, we have $S = -\sigma J_{S^*}^{-1} v$ and therefore $\Sigma = \langle S, S^T \rangle = \sigma^2 J_{S^*}^{-1} J_{S^*}^{-T}$. This approximation assumes that the instant change of activity remains near zero at a relatively low level of noise strength $\sigma$ [49], providing a clear mapping relationship between the block matrices within the covariance $\Sigma$ and the full Jacobian $J_{S^*}$, then This can be derived from the following equation which provides a direction expression of how asymmetric Jacobian shapes steady covariance $\Sigma$ of the system [49]:

$$\Sigma = \sigma^2 J_{S^*}^{-1} J_{S^*}^{-T}, \tag{45}$$

where $\Sigma = \begin{bmatrix} COV_{EE} & COV_{EI} \\ COV_{IE} & COV_{II} \end{bmatrix}$ and with the knowledge of Schur Complement [68], we have:

$$J_{S^*}^{-1} = \begin{bmatrix} P_{EE} & -P_{EE} J_{EI} J_{II}^{-1} \\ -J_{II}^{-1} J_{IE} P_{EE} & J_{II}^{-1} + J_{II}^{-1} J_{IE} P_{EE} J_{EI} J_{II}^{-1} \end{bmatrix}$$

where we have $P_{EE} = (J_{EE} - J_{EI} J_{II}^{-1} J_{IE})^{-1}$, and this implies that

$$COV_{EE} = \sigma^2 P_{EE} P_{EE}^T + \sigma^2 P_{EE} J_{EI}^2 J_{II}^{-2} P_{EE}^T,$$

$$COV_{IE} = -\sigma^2 J_{II}^{-1} J_{IE} P_{EE} P_{EE}^T - \sigma^2 J_{II}^{-1} J_{IE} P_{EE} J_{EI}^2 J_{II}^{-2} P_{EE}^T - \sigma^2 J_{II}^{-2} J_{EI} P_{EE}^T$$

$$\approx -\sigma^2 J_{II}^{-1} J_{IE} P_{EE} P_{EE}^T - \sigma^2 J_{II}^{-1} J_{IE} P_{EE} J_{EI}^2 J_{II}^{-2} P_{EE}^T.$$

Therefore, we can solve that:

$$J_{EI} COV_{IE} COV_{EE}^{-1} \approx -J_{EI} J_{II}^{-1} J_{IE}. \tag{46}$$

Taking Eq.46 in Eq.44, we finally obtain the ground truth EC $J_{eff}$ in Eq. 11.

The ground truth elements of EC $J_{eff}$ is:

$$J_{eff} = \begin{cases} -\dfrac{1}{\tau_E} - \gamma H_E^{i*} + \gamma\left(1 - S_E^{i*}\right)\left(w_{EE} g_E^i - \dfrac{w_{EI} w_{IE} g_E^i g_I^i}{\dfrac{1}{\tau_I} + g_I^i}\right), i = j \\ \gamma G\left(1 - S_E^{i*}\right) g_E^i C_{ij}, i \neq j \end{cases}, \tag{47}$$

where $H_E^{i*} = H_E^i(x_E^{i*})$, and $g_P^i = \dfrac{dH_P^i}{dx_P^i}\big|_{x_P^{i*}}$ is the firing rate change (gain function level).

For comparison, the full Jacobian matrix $J_{S^*}$ is:

$$J_{S^*} = \begin{bmatrix} -\dfrac{1}{\tau_E} - \gamma H_E^{i*} + \gamma w_{EE}\left(1 - S_E^{i*}\right) g_E^i, i = j & -\gamma w_{EI}\left(1 - S_E^{i*}\right) g_E^i, i = j \\ \gamma G\left(1 - S_E^{i*}\right) g_E^i C_{ij}, i \neq j & 0, i \neq j \\ w_{IE} g_I^i, i = j & -\dfrac{1}{\tau_I} - g_I^i, i = j \\ 0, i \neq j & 0, i \neq j \end{bmatrix}. \tag{48}$$

## 2.H Sampling Interval Effect on EC reconstruction.

In this section, we derive how sampling intervals affect EC estimation using Volterra expansion [67]. The Volterra expansion quantifies the impact of perturbations on nonlinear systems by expanding system responses before and after perturbation, where here noise serves as continuous perturbation influencing the neural dynamics. We start with the noise-driven nonlinear Model A (Eqs. 1-3) and write it into a general nonlinear dynamic:

$$\frac{dx}{dt} = f(x;\theta) + \sigma v(t),$$

where $\theta$ represents system parameter and $v(t)$ is $i.i.d.$ white noise. Apply the first-order Volterra expansion to the activity $x(t)$ at $t_0$ [67], we have:

$$x(t) = x(t_0) + \int_{t_0}^{t} f(x;\theta)dt' + \sigma \int_{t_0}^{t} K(t,t')v(t')dt' + O(\sigma^2), \tag{50}$$

where the first term is the initial state at $t_0$, the second term is the deterministic evolution (or zero-order Volterra kernel, representing the behavior of the dynamics when there is no outer input), $K(t,t') = \frac{\delta x(t)}{\delta v(t')}$ is the linear response function (Green's function of Model A, or the first-order Volterra kernel) and $\delta(\cdot)$ denotes functional derivative [5]. The last term $O(\sigma^2)$ denotes higher-order kernel with respect to the noise strength $\sigma$.

To have an explicit approximated solution of EC, we linearized this general nonlinear dynamic near its stable state $x^*$ and we have:

$$\frac{dx}{dt} = J(x - x^*) + \sigma v(t).$$

We can calculate the linear response function according to the definition of Green's function [45]: $\left(\frac{d}{dt} - J\right) K(t,t') = \delta(t - t')$, this implies $K(t,t') = e^{J(t-t')}$.

We then pick $t = t_0 + T$ represents the next time point after sampling at frequency $\frac{1}{T}$ (where $T = nh + \Delta \approx nh$, i.e., sampling activity every $n$ time steps, $0 < \Delta < h$) and take the linear approximation into Eq. 50:

$$x(t_0 + T) \approx x^* + e^{JT}[x(t_0) - x^*] + \int_{t_0}^{t_0+T} e^{J(T+t_0-t)} \nu(t')dt'$$

$$= x^* + e^{JT}[x(t_0) - x^*] + \int_0^T e^{J(T-s)} \nu(s)ds$$

$$\approx x^* + e^{JT}[x(t_0) - x^*] + B_T \nu(t_0 + T), \qquad (51)$$

where in the last line, we simplify the white noise integral $\int_0^T e^{J(T-s)} \nu(s)ds$ into a mapping matrix $B_T$, where $B_T B_T^T = \sigma^2 \int_0^T e^{Js} e^{J^T s}ds$ [44]. Note that $\nu(t)$ represents Gaussian white noise with $\delta$-correlation, the integral should be properly interpreted as an Itô stochastic integral with respect to a Wiener process. For brevity and since we consider only additive noise, we maintain the informal notation throughout [45]. Thus, we have derived how we naturally approximate the neural activity during the time interval $[t_0, t_0 + T]$ [44].

We then use Eq. 51 to derive how this sampled neural activity contributes to EC estimation. We first minus $x(t_0)$ and devide by $T$ at both sides:

$$\frac{x(t_0 + T) - x(t_0)}{T} = \frac{e^{JT} - I}{T}[x(t_0) - x^*] + \frac{B_T}{T} \nu(t_0 + T),$$

The ground truth EC at sampling interval $T$ is calculated:

$$J_T = < \frac{x(t_0 + T) - x(t_0)}{T}, x(t_0) > < x(t_0), x(t_0) >^{-1}$$

$$= < \frac{e^{JT} - I}{T} [x(t_0) - x^*] + \frac{B_T}{T} \nu(t_0 + T), x(t_0) > < x(t_0), x(t_0) >^{-1}$$

$$= \left[ < \frac{e^{JT} - I}{T} [x(t_0) - x^*], x(t_0) > + < \frac{B_T}{T} \nu(t_0 + T), x(t_0) > \right] < x(t_0), x(t_0) >^{-1}$$

$$= < \frac{e^{JT} - I}{T} [x(t_0) - x^*], x(t_0) > < x(t_0), x(t_0) >^{-1}$$

$$= < \frac{e^{JT} - I}{T} x(t_0), x(t_0) > < x(t_0), x(t_0) >^{-1}$$

$$= \frac{e^{JT} - I}{T}, \tag{52}$$

where in the third line, the covariance $< \frac{B_T}{T} \nu(t_0 + T), x(t_0) > = 0$ because of the independence of Gaussian white noise.

To estimate the true Jacobian matrix $\hat{J}_o$ from the reconstructed $\hat{J}_T$, we solved Eq.16 as follows: rearrange Eq.16 and take the matrix logarithm of both sides,

$$T J_T = e^{T J_o} - I,$$

$$\ln(T J_T + I) = T J_o,$$

then dividing by $T$ we have isolated the $J_o$:

$$J_o = \frac{\ln(T J_T + I)}{T}.$$

Substituting the estimated $\hat{J}_T$ using sampled neural activity in Eq. 17, we have the estimation of true EC in Eq. 18.

# Acknowledgments

This work was partially supported by Science and Technology Innovation 2030-Major Projects (No. 2022ZD0208500 to Changsong Zhou), the Hong Kong Research Grant Council Senior Research Fellow Scheme (SRFS2324-2S05 to Changsong Zhou) and General Competitive Fund (GRF12202124 to Changsong Zhou). This research was conducted using the resources of the High Performance Computing Cluster Centre, HKBU, which receives funding from the RGC, University Grant Committee of Hong Kong and HKBU. The funders had no role in study design, data collection and analysis, decision to publish, or preparation of the manuscript.

# Competing Interests

The authors have declared that no competing interests exist.

# Author Contributions

**Conceptualization:** Jiawen Chang, Zhuda Yang, Changsong Zhou.

**Formal analysis:** Jiawen Chang.

**Investigation:** Jiawen Chang, Zhuda Yang.

**Methodology:** Jiawen Chang.

20 **Supervision:** Changsong Zhou.

21 **Writing – original draft:** Jiawen Chang.

22 **Writing – review & editing:** Jiawen Chang, Zhuda Yang, Changsong Zhou

# Supporting Information

**S1. Supplementary Methods**

    **S1.1 Resimulation of Model A**

    **S1.2 Mapping of Heterogeneity Pairs – Model A and Model B**

    **S1.3 Implement of Feedback Inhibition Control**

    **S1.4 Resimulation of Model C**

**S2. Supplementary Figures**

    **Figure S1: Robustness of reconstruction in effective heterogeneity and asymmetry.**

242    **Figure S2: Reconstruction performance in effective heterogeneity**
243    **and asymmetry with/without ground truth properties.**

244    **Figure S3: Robustness of reconstruction in effective heterogeneity**
245    **and asymmetry using alternative EC estimations.**

246    **Figure S4: Detailed characterization of model performance and**
247    **parameter estimation reliability.**

248    **Figure S5: Robustness of reconstruction in FC and timescale in**
249    **mapping heterogeneous parameters.**

250    **Figure S6: Robustness of reconstruction in effective heterogeneity**
251    **and asymmetry using E-I model.**

252    **Figure S7: Diagonal analysis of Large-scale Balanced Network with**
253    **Localized Inhibitory.**